\documentclass[12pt]{article}
\usepackage{graphics,color}
\begin{document}
\thispagestyle{empty}
\noindent\
\\
\\
\\
\begin{center}
\large \bf Neutrino Masses and Flavor Mixing 
\end{center}
\hfill
 \vspace*{1cm}
\noindent
\begin{center}
{\bf Harald Fritzsch}\\
Department f\"ur Physik, Universit\"at M\"unchen,\\
Theresienstra{\ss}e 37, 80333 M\"unchen
\vspace*{0.5cm}
\end{center}

\begin{abstract}

We discuss the neutrino oscillations, 
using texture zero mass matrices for the leptons. The reactor mixing angle $\theta^{}_{l}$ is calculated. The ratio of the masses of two neutrinos is determined by the solar mixing angle. We can calculate the masses of the three neutrinos:

$m_1$ $\approx$ 0.003 eV - $m_2$ $\approx$ 0.012 eV - $m_3$ $\approx$ 0.048 eV. 
\end{abstract}

\newpage

The flavor mixing of the quarks is parametrized by the CKM-matrix. 
There are several ways to describe the CKM-matrix in terms of three 
angles and one phase parameter. I prefer the parametrization, given below, 
which Z. Xing and I introduced years ago (ref.(1)), given by the angles $\theta_u$,  
$\theta_d$ and $\theta$:
\begin{eqnarray}
U = \left( \matrix{ c^{}_{u}
& s^{}_{u} & 0 \cr
-s^{}_{u} &
c^{}_{u} &
0 \cr 0 & 0 & 1 \cr} \right)\times \left( \matrix{ e^{-i\delta}
& 0 & 0 \cr
0 &
c &
s \cr 0 & -s & c \cr} \right) \times \left( \matrix{ c^{}_{d}
& - s^{}_{d} & 0 \cr
s^{}_{d} &
c^{}_{d} &
0 \cr 0 & 0 & 1 \cr} \right),
\end{eqnarray}
where $c^{}_{u,d} \sim \cos\theta^{}_{u,d}$, $s^{~}_{u,d}
\sim \sin\theta^{}_{u,d}$, $c \sim \cos\theta$ and $s \sim
\sin\theta$.\\

The angle $\theta_u$ describes the mixing between the quarks "u - c", the angle $\theta_d$ the mixing between the quarks "d - s" and the angle $\theta$ the mixing among the heavy quarks "t,c - b,s". The three angles have been determined by the experiments:\\ 

$\theta^{}_u \simeq 5.4^\circ$, ~~~~~  $\theta^{}_d \simeq 11.7^\circ$, $ ~~~~~ \theta^{}\simeq 2.4^\circ$.\\
\\
Presumably the flavor mixing angles are not fixed values, but functions of the quark masses. If the masses change, the mixing angles will also change. For example, the Cabibbo angle 
$\theta^{}_C \simeq 13^\circ$ could be given by the ratio of the quark masses: \\
\begin{equation}
\tan\theta^{}_C \simeq
\sqrt{m^{}_d/m^{}_s}.
\end{equation}
\\
This relation works very well: 
\begin{equation}
\tan\theta^{}_C \simeq 0.23,
\end{equation}
\begin{equation}
\sqrt{m^{}_d/m^{}_s}\simeq 0.23.
\end{equation}
\\
Such a relation can be derived, if the quark mass matrices have "texture zeros", as shown by S. Weinberg and me in 1977 (ref.(2)).\\ 

Let me discuss a simple example, using only four quarks: u,d - c,s. Their mass matrices have a zero in the (1,1)-position: 
\begin{eqnarray}
M= \left( \matrix{ 0 & A \cr A^* & B \cr
} \right)\;.
\end{eqnarray}

These mass matrices can be diagonalized by a rotation. The rotation angles are: 
\begin{equation}
\theta^{}_u \simeq
\sqrt{m^{}_u/m^{}_c}\simeq 0.09,\hspace*{1cm}
\theta^{}_d \simeq
\sqrt{m^{}_d/m^{}_s}\simeq 0.23.
\end{equation}

The Cabibbo angle is given by the difference:
\begin{equation}
\theta^{}_C \simeq
|\sqrt{m^{}_d/m^{}_s}-e^{i\phi}\sqrt{m^{}_u/m^{}_c}|.
\end{equation}
In the complex plane this relation describes a triangle. The phase parameter is unknown, however it must be close to 90 degrees, since the Cabibbo angle is given by the ratio $m_d$/$m_s$:
\begin{equation}
\theta^{}_C \simeq
\sqrt{m^{}_d/m^{}_s}.
\end{equation}
Thus the triangle is close to a rectangular triangle.\\ 

For six quarks the "texture zero" mass matrices for the quarks of charge (2/3) and of charge (-1/3) are: 
\begin{eqnarray}
M= \left( \matrix{ 0 & A & 0 \cr A^* & B & C \cr
0 & C^* & D \cr} \right) \;.
\end{eqnarray}
\\
We can calculate the angles $\theta_u$ and   
$\theta_d$ as functions of the mass eigenvalues: 
\begin{equation}
\theta^{}_u \simeq
\sqrt{m^{}_u/m^{}_c},\hspace*{1cm}
\theta^{}_d \simeq
\sqrt{m^{}_d/m^{}_s}.
\end{equation}

Using the observed mass values for the quarks, we find:\\ 

$\theta^{}_d \simeq (13.0\pm 0.4)^\circ,\hspace*{1cm} \theta^{}_u \simeq (5.0\pm 0.7)^\circ.$\\ 
\\
The experimental values agree very well with the theoretical results:\\

~~~~~~~~~~~~~~~~$\theta^{}_d \simeq (11.7\pm 2.6)^\circ,\hspace*{1cm} \theta^{}_u \simeq (5.4\pm 1.1)^\circ$.\\
\\
Now we consider the flavor mixing of the leptons. The neutrinos, emitted in weak decays, are mixtures of 
different mass eigenstates. This leads to neutrino 
oscillations - at least two neutrinos must have finite masses.

The lepton flavor
mixing is described by a $3\times 3$ unitary matrix $U$, analogous to the CKM mixing matrix
for the quarks. It can be parametrized in terms of three angles and three phases. I use 
a parametrization, introduced by Z. Xing and me (ref.(3)):
\begin{eqnarray}
U = \left ( \matrix{ s^{}_l s^{}_{\nu} c + c^{}_l c^{}_{\nu}
e^{-i\varphi} & s^{}_l c^{}_{\nu} c - c^{}_l s^{}_{\nu}
e^{-i\varphi} & s^{}_l s \cr c^{}_l s^{}_{\nu} c - s^{}_l c^{}_{\nu}
e^{-i\varphi} & c^{}_l c^{}_{\nu} c + s^{}_l s^{}_{\nu}
e^{-i\varphi} & c^{}_l s \cr - s^{}_{\nu} s   & - c^{}_{\nu} s   & c
\cr } \right ) P^{}_\nu \; ,   
\end{eqnarray}
where $c^{}_{l,\nu} \sim \cos\theta^{}_{l,\nu}$, $s^{~}_{l,\nu}
\sim \sin\theta^{}_{l,\nu}$, $c \sim \cos\theta$ and $s \sim
\sin\theta$. The angle $\theta^{}_{\nu}$ is the solar angle $\theta^{}_{sun}$, the angle 
$\theta$ is the atmospheric angle $\theta^{}_{at}$, and the angle 
$\theta^{}_{l}$ is the "reactor angle". The phase matrix $P^{}_\nu ={\rm
Diag}\{e^{i\rho}, e^{i\sigma}, 1\}$ is relevant only, if
the neutrino masses are Majorana masses.\\

The neutrino oscillations are described by two large mixing angles:\\

$\theta^{}_{sun}=\theta^{}_\nu \simeq 34^\circ$, \hspace*{1cm} 
$\theta^{}_{at}=\theta\simeq 45^\circ $.\ 
\\
\\
The reactor angle $\theta^{}_{l}$ is much smaller: $\theta^{}_l \simeq 13^\circ $.\
\\

We assume that the mass matrices of the leptons have "texture zeros":
\begin{eqnarray}
M= \left( \matrix{ 0 & A & 0 \cr A^* & B & C \cr
0 & C^* & D \cr} \right) \;.
\end{eqnarray}\\
In this case we can calculate two leptonic mixing angles as functions of mass ratios: 
 
\begin{equation}
\tan\theta^{}_l \simeq
\sqrt{m^{}_e/m^{}_\mu}, \hspace*{1cm}
\tan\theta^{}_\nu \simeq
\sqrt{m^{}_1/m^{}_2}.
\end{equation}
\\
From the solar mixing angle we obtain for the neutrino mass ratio:
\begin{equation}
{m^{}_1/m^{}_2} \simeq 0.42.
\end{equation}

This relation and the experimental results for the mass differences of the neutrinos, measured 
by the neutrino oscillations, allow us to 
determine the three neutrino masses:

\begin{eqnarray}
{m^{}_1} \simeq 0.003~eV , \nonumber \\
{m^{}_2} \simeq 0.012~eV , \nonumber \\
{m^{}_3} \simeq 0.048~eV.
\end{eqnarray}

We expect that the mass matrices of the quarks and leptons are not exactly given by 
texture zero matrices. Radiative corrections of the 
order of the fine-structure constant $\alpha$ will contribute - the zeros will be replaced by small numbers.\\ 

The ratios of the masses of the quarks with the same electric charge 
seem to be universal:
 
\begin{equation}
 \begin{array}{c}
\frac{m^{}_u}{m^{}_c} \simeq \frac{m^{}_c}{m^{}_t} \simeq 0.005, \\
\\
\frac{m^{}_d}{m^{}_s} \simeq \frac{m^{}_s}{m^{}_b} \simeq 0.044.
\end{array}
\end{equation}
The dynamical reason for this universality is unclear. It might follow 
from specific properties of the texture zero mass matrices.  But in the 
case of the charged leptons there is no universality:
 \begin{equation}
 \begin{array}{c} 
\frac{m^{}_e}{m^{}_\mu} \simeq 0.005, \\
\\
\frac{m^{}_\mu}{m^{}_\tau} \simeq 0.06.
\end{array}
\end{equation}
If the ratios of the charged lepton masses were universal, the mass of the electron would have 
to be about 6 MeV.\\ 

The universality is not expected to 
be exact, due to radiative corrections.  
A radiative correction of the order of $ \pm (\alpha/\pi){m^{}_\tau} \simeq \pm 4~MeV $ would have 
to be added to the charged 
lepton masses. Such a contribution is relatively small for the muon and the tauon, but large
in comparison to the observed electron mass. One expects that the physical electron mass 
is the sum of a bare electron mass $M^{}_e$, due to the texture zero mass matrix, and a radiative 
correction $R^{}_e$:\\

 ${m^{}_e} = M^{}_e - R^{}_e $.\\ 
\\
If we assume, that the bare electron mass is given by the universality, we obtain:\\
 
$M^{}_e$= 5.51 MeV, $R^{}_e$=5.00 MeV.\\ 
\\
Radiative corrections also contribute to the muon and the tauon mass, but 
here the corrections are small in comparison to the bare masses and will be neglected.\\

We calculate the angle $\theta^{}_l $ ( see eq. (13)): 
\begin{eqnarray}
\tan\theta^{}_l \simeq
\sqrt{M^{}_e/m^{}_\mu}\simeq 0.23, \nonumber \\
\theta^{}_{l}\simeq 13^\circ. 
\end{eqnarray}
This angle agrees with the experimental result.\\ 

If we would not have taken into account 
the radiative corrections for the electron mass, we would obtain a value for the reactor angle, which is much smaller than the experimental value: 
\begin{eqnarray}
\tan\theta^{}_l \simeq
\sqrt{m^{}_e/m^{}_\mu}\simeq 0.07, \nonumber \\
\theta^{}_{l}\simeq 4^\circ.
\end{eqnarray}

The neutrino masses, given in eq. (15), are very small, much smaller than the masses of the charged leptons. The ratio of the mass of the tau neutrino and of the mass of the tauon is only about 
$2.7\times 10^{-11} $.\\ 

Are the neutrino masses normal Dirac masses, as the masses of the charged 
leptons and quarks, or are they Majorana masses? In this case the smallness of the neutrino masses can be understood by the "seesaw"-mechanism (ref.(5)). Here the neutrino masses are mixtures of Majorana and Dirac masses.\\ 

The mass matrix of the neutrinos is a matrix with one "texture zero" in the (1,1)-position. The two off-diagonal terms are given by the Dirac mass term D - a large Majorana mass term appears in the (2,2)-position:
 
\begin{eqnarray}
M_\nu = \left( \matrix{ 0 & D \cr D & M \cr
} \right)\;.
\end{eqnarray}
\\
After diagonalization one obtains a large Majorana mass M and a small neutrino mass: 

\begin{equation}
  {m^{}_\nu}\simeq D^{2}/M. 
\end{equation}
\\
One expects that the Dirac term D is similar to the corresponding charged lepton mass. As an example we 
consider a Dirac mass matrix D with the eigenvalues: 
\begin{eqnarray}
D= \left( \matrix{ {M^{}_1} & 0 & 0 \cr 0& {M^{}_2} & 0 \cr
0 & 0 & {M^{}_3} \cr} \right) \;.
\end{eqnarray}

We assume that these Dirac masses are proportional to the neutrino masses, given in eq. (15). As an example we consider the mass values: \\
 
$ M_1 \simeq 30~ MeV, ~~M_2 \simeq 120~ MeV, ~~M_3 \simeq 480~ MeV $.\\ 
\\
For the heavy Majorana mass M we find in this case:\\

$~~~~~~~~~~~~~~~~~~~~~~~~~~~~~~~~~~~~~M = 4.8 \times 10^9 GeV$.\\

The only way to test the nature of the neutrino masses is to study the neutrinoless double beta decay. Some heavy nuclei decay by double beta decay - two neutrons emit electrons and antielectron-neutrinos, e.g. the decay of selenium into krypton, which is a normal double beta decay.\\ 

If neutrinos are Majorana particles, the two antielectron-neutrinos can annihilate. Only two electrons are emitted - this is the neutrinoless double beta decay, which violates lepton number conservation.\\

Thus far the neutrinoless double beta decay has not been observed. For example, in the Cuoricino experiment in the Grand Sasso Laboratory one searches for the neutrinoless double beta decay of tellurium.\\ 

If neutrinos mix, all three neutrino masses will contribute to the decay rate for neutrinoless double 
beta decay. Their contributions are given by the mass of the neutrino, multiplied by the square of the 
transition element, including the CP-violation phases.\\ 

Using the neutrino masses, given in eq. (15) and the transition elements, given by the mixing angles, I find for the effective neutrino mass, relevant for 
the neutrinoless double beta decay:

\begin{equation}
  \widetilde{m}\simeq 0.016~~ eV.
\end{equation}
The present limit for this effective mass is provided by the Cuoricino experiment: $\widetilde{m} < 0.23 ~~eV $.\\  

The texture zero idea provides a coherent framework to understand the flavor mixing of quarks and leptons. Two of the three mixing angles for the quarks are given by quark mass ratios. For the leptons the solar mixing angle 
is determined by the ratio of the masses of two neutrinos. Thus the absolute masses of the neutrinos are fixed and very small. The reactor angle is given by the ratio of the bare electron mass and the muon mass.

\end{document}